\documentclass[slac_one]{revtex4}
\usepackage{graphicx}
\usepackage{fancyhdr}
\newcommand{\bbbar}{\ensuremath{B\overline{B}} }
\newcommand{\btaunu}{\ensuremath{B^{-}\rightarrow\tau^{-}\overline{\nu}}}
\newcommand{\btag}{\ensuremath{B_{\rm tag}} }
\newcommand{\bsig}{\ensuremath{B_{\rm sig}} }
\newcommand{\eecl}{\ensuremath{E_{\text{ECL}}} }
\newcommand{\brvalue}{1.65}
\newcommand{\brstaterr}{^{+0.38}_{-0.37}}
\newcommand{\brsysterr}{^{+0.35}_{-0.37}}

\begin{document}

\title{Rare B decays with leptons at Belle} 

%

\author{K. Hara for the Belle Collaboration}
\affiliation{Nagoya University, Furou-cho Chikusa-ku Nagoya, Aichi, 464-8602, Japan}

\begin{abstract}
We present a new measurement of the purely leptonic decay 
$B^{-}\rightarrow\tau^{-}\overline{\nu}_\tau$ with a semileptonic $B$ tagging method,
using a data sample containing $657\times 10^6$ \bbbar pairs collected 
with the Belle detector at the KEKB asymmetric $e^{+}e^{-}$ collider.
A sample of $\bbbar$ pairs are tagged by reconstructing one $B$ meson decaying
semileptonically. We detect the $B^-\to \tau^-\overline{\nu}_{\tau}$ candidate in
the recoil.
We obtain a signal with a significance of 3.8 standard deviations including systematics,
and measure the branching fraction to be
${\cal B}(B^{-}\rightarrow\tau^{-}\overline{\nu}_{\tau}) = 
 (\brvalue\brstaterr(\text{stat})\brsysterr(\text{syst})) \times 10^{-4}$.
This result confirms the evidence for $B^{-}\to\tau^-\overline{\nu}_\tau$ 
obtained in the previous Belle measurement with a hadronic $B$ tagging method.
The $B$ meson decay constant $f_B$ and constraint on charged Higgs are 
obtained using the measured branching fraction.
\end{abstract}

\maketitle

\thispagestyle{fancy}

\section{Introduction}
The purely leptonic decay $B^{-}\rightarrow\tau^{-}\overline{\nu}$~\cite{conjugate}
is of particular interest since it provides a direct measurement of the product of 
the Cabibbo-Kobayashi-Maskawa(CKM) matrix element $V_{ub}$~\cite{CKM} and
the $B$ meson decay constant $f_B$. In the Standard Model(SM), 
the branching fraction of the decay $B^{-}\rightarrow\tau^{-}\overline{\nu}$ 
is given by
\begin{equation}
 \label{eq:BR_B_taunu}
{\cal B}(B^{-}\rightarrow\tau^{-}\overline{\nu})_{SM} = \frac{G_{F}^{2}m_{B}m_{\tau}^{2}}{8\pi}\left(1-\frac{m_{\tau}^{2}}{m_{B}^{2}}\right)^{2}f_{B}^{2}|V_{ub}|^{2}\tau_{B},
\end{equation}
where $G_{F}$ is the Fermi coupling constant, $m_{\tau}$ and $m_{B}$ are
the $\tau$ lepton and $B$ meson masses, and $\tau_{B}$ is the $B^{-}$ lifetime.
Physics beyond the SM, such as supersymmetry or two-Higgs doublet models, could suppress
or enhance ${\cal B}(B^{-}\rightarrow\tau^{-}\overline{\nu})$ to levels several 
times as large as the SM expectation through 
the introduction of a charged Higgs boson \cite{Hou:1992sy,Baek:1999ch}.
The charged Higgs effect is described as 
\begin{eqnarray}
    \label{eq:BR_higgs}
    {\cal B}(B^{-}\rightarrow\tau^{-}\overline{\nu}) &=& {\cal B}(B^{-}\rightarrow\tau^{-}\overline{\nu})_{SM} \times r_H, \nonumber \\
    r_H &=&(1-\frac{M_B^2}{m_{H^\pm}^2} \tan^2 \beta )^2,
\end{eqnarray}
where $m_H^{\pm}$ is the charged Higgs mass and $\tan\beta$ is the ratio of
the two Higgs vacuum expectation values.
The expected SM branching fraction from other experimental constraints is
$(0.78^{+0.09}_{-0.13})\times 10^{-4}$~\cite{CKMfitter2008}.
No statistically significant enhancement relative to the SM expectation has been observed
in previous experimental studies.
The previous Belle measurement\cite{ikado-2006-97} reported the first evidence of 
 $B^{-}\rightarrow\tau^{-}\overline{\nu}$ decay
 with a significance of $3.5$ standard deviations ($\sigma$),
 and measured the branching fraction to be
${\cal B}(B^{-}\rightarrow\tau^{-}\overline{\nu}_{\tau}) = 
 (1.79^{+0.56}_{-0.49}(\mbox{stat})^{+0.46}_{-0.51}(\mbox{syst})) \times 10^{-4}$, 
 using a full reconstruction tagging method.
The BaBar Collaboration has reported a search for $B^{-}\rightarrow\tau^{-}\overline{\nu}$
decay with hadronic tagging~\cite{Aubert:2007} and
semileptonic tagging~\cite{Aubert:2007_semil} using $383 \times 10^6$ \bbbar pairs. 
They report a 2.6 $\sigma$ excess, combining the two measurements.
To establish the $\btaunu$ signal, test consistency with the SM and search for a charged 
Higgs boson effect, we need more statistics.

In this report, we present a new measurement of $B^{-}\rightarrow\tau^{-}\overline{\nu}_{\tau}$
from the Belle experiment with a semileptonic tagging method, 
based on a $605~\textrm{fb}^{-1}$ data sample containing 
$657\times 10^{6}$ \bbbar pairs collected with the Belle detector~\cite{belle-detector}
at the KEKB asymmetric energy $e^{+}e^{-}$ ($3.5$ on $8$ GeV) collider~\cite{KEKB}
operating at the $\Upsilon(4S)$ resonance ($\sqrt{s} = 10.58$ GeV).

\section{Measurement of $B^{-}\rightarrow\tau^{-}\overline{\nu}_\tau$ with a semileptonic $B$ tagging method}
The strategy adopted for this analysis is same as in the previous measurements.
We reconstruct one of the $B$ mesons decaying semileptonically (referred to hereafter as $\btag$)
and compare the properties of the remaining particle(s) in the event ($\bsig$) to those
expected for signal and background.
We reconstruct the \btag in $B^-\to D^{*0}\ell^-\overline{\nu}$ and
$B^-\to D^{0}\ell^-\overline{\nu}$ decays.
For $D^{*0}$ reconstruction, we use $D^{*0}\to D^{0}\pi^0$ and $D^{0}\gamma$ decays.
$D^0$ mesons are reconstructed in $K^-\pi^+$, $K^-\pi^+\pi^0$ and $K^-\pi^+\pi^-\pi^+$.
For $\bsig$, we use $\tau^-$ decays to 
$\tau^- \to \ell^- \overline{\nu}_\ell \nu_\tau$, where $\ell = \mu$ or $e$, and
$\tau^- \to \pi^- \nu_\tau$.
We require that no charged particle or $\pi^0$ remain in the event after removing
the particles from the \btag and \bsig candidates.
The selection criteria for \btag and \bsig are optimized for each of the $\tau$ decay modes,
because the background levels and the background components are mode-dependent.
The details of the selection criteria are described elsewhere~\cite{belle-conf0840}.

\begin{figure}[h]
    \begin{center}
	\includegraphics[width=0.6\textwidth]{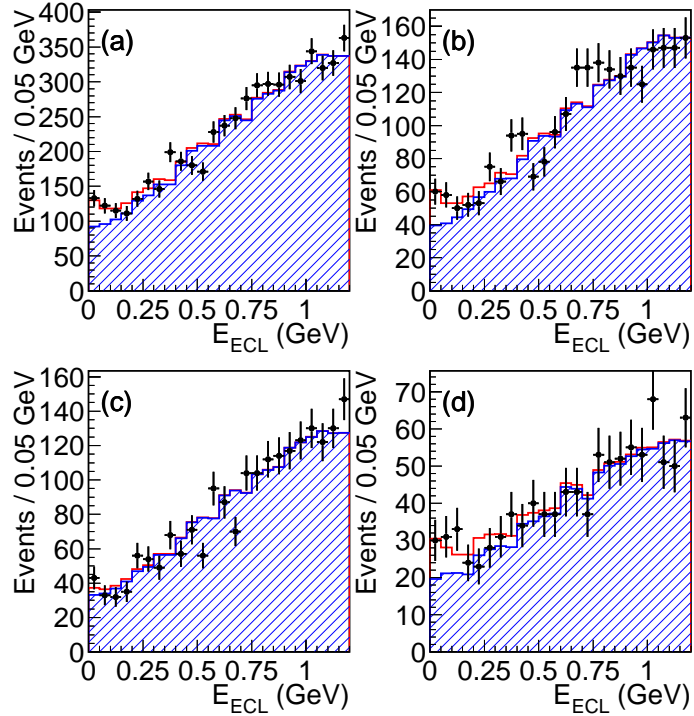}
	\caption{\eecl distribution of semileptonic tagged events 
	  with the fit result for 
	  (a) all $\tau$ decay modes combined,
	  (b) $\tau^- \to e^-\overline{\nu}_e\nu_\tau$,
	  (c) $\tau^- \to \mu^-\overline{\nu}_\mu\nu_\tau$ and
	  (d) $\tau^- \to \pi^-\nu_\tau$.
	  The points with error bars are data. The hatched histogram
	  and solid open histogram are the background and the signal, 
	  respectively.
	  }
	\label{fig:eecl_fit}
    \end{center}
\end{figure}

The most powerful variable for separating signal and background is the
remaining energy in the electromagnetic calorimeter (ECL), denoted $\eecl$, which is the sum of
the energies of ECL clusters that are not associated with 
particles from the \btag and \bsig candidates.
The number of signal events is extracted from an extended maximum likelihood fit to the \eecl
distribution.
We combine $\tau$ decay modes by constraining the ratios of the signal 
yields to the ratio of reconstruction efficiencies obtained from MC.
Figure~\ref{fig:eecl_fit} shows the $E_{\rm ECL}$ distribution with the fit results.
We see a clear excess of signal events in the region near $E_{\rm ECL}\sim 0$.
Table~\ref{tab:fit_result} summarizes the signal yields and the branching fractions
obtained from separate fits for each $\tau$ decay mode and fits with all three modes combined.
\begin{table}[h]
\begin{tabular}{lcccc} \hline \hline
Decay Mode & Signal Yield   &  $\varepsilon$ & $\cal B$  \\ \hline
$\tau^-\to e^{-}\nu\bar{\nu}_{\tau}$       &$78^{+23}_{-22}$  & $5.9\times 10^{-4}$ & $(2.02^{+0.59}_{-0.56})\times 10^{-4}$ \\
$\tau^-\to \mu^{-}\nu\bar{\nu}_{\tau}$     &$15^{+18}_{-17}$  & $3.7\times 10^{-4}$ & $(0.62^{+0.76}_{-0.71})\times 10^{-4}$ \\
$\tau^-\to\pi^{-}\nu_{\tau}$               &$58^{+21}_{-20}$  & $4.7\times 10^{-4}$ & $(1.88^{+0.70}_{-0.66})\times 10^{-4}$ \\
\hline
Combined                                   &$154^{+36}_{-35}$ & $14.3\times 10^{-4}$ & $(1.65^{+0.38}_{-0.37})\times 10^{-4}$ \\
\hline\hline
    \end{tabular}
    \caption{
      Results of the fit for signal yields and branching fractions.
      Errors are statistical only.
    }
   \label{tab:fit_result}
\end{table}
Systematic errors for the measured branching fraction are associated with 
the uncertainties in the signal yield, efficiencies and number of $B^{+}B^{-}$ pairs. 
The systematic errors for the signal yield arise from the uncertainties in the PDF shapes
for the signal ($^{+3.1}_{-3.2}$\%) and for the background ($^{+11.8}_{-11.2}$\%)
which are dominated by MC statistics. 
For the latter, uncertainties in the branching fractions of $B$ decay modes that 
peak at $\eecl=0$ such as $B^-\to D^0\ell^+\nu$ with $D^0\to K_L^0 \pi^0, K_L^0 K_L^0$
and so on ($^{+4.2}_{-8.4}$)\%, 
as well as uncertainties in the background from rare $B$ decays and
$\tau$ pair events ($3.8$\%) are also taken into account.
We take a 11.6\% error as the systematic error associated with the tag reconstruction 
efficiency from the difference of yields between data and MC for the control sample.
This value includes the error in the branching fraction
${\cal B}(B^{-}\rightarrow D^{*0}\ell^{-}\bar{\nu})$, which we estimate from 
${\cal B}(B^{0}\rightarrow D^{*-}\ell^{+}\nu)$ in Ref.~\cite{pdg2006}
and isospin symmetry.
The systematic error in the signal efficiencies arises from the uncertainty in tracking
efficiency (1.0\%), particle identification efficiency (1.3\%),
branching fractions of $\tau$ decays (0.4\%), and MC statistics (0.9\%).
The systematic error due to the uncertainty in the number of $\bbbar$ pairs is 1.4\%.
The total fractional systematic uncertainty is $^{+21}_{-22}\%$.
We obtain the branching fraction to be 
\begin{equation}
{\cal B}(B^{-}\rightarrow\tau^{-}\bar{\nu}_{\tau}) = (\brvalue\brstaterr(\text{stat})\brsysterr(\text{syst}))\times 10^{-4}.
\end{equation}
The significance of the observed signal is estimated to be 3.8 $\sigma$ including
systematic errors.

\section{Determination of $f_B$ and Constraint on Charged Higgs}
Using the measured branching fraction and known values of $G_F$, $m_B$, 
$m_{\tau}$ and $\tau_B$~\cite{pdg2006}, the product
of the $B$ meson decay constant $f_B$ and the magnitude of the
Cabibbo-Kobayashi-Maskawa matrix element $|V_{ub}|$ is determined to be 
$f_B |V_{ub}|= (9.7\pm1.1^{+1.0}_{-1.1}) \times 10^{-4}$ GeV.
Combining it with $|V_{ub}|=3.99^{+0.35}_{-0.30}$ by HFAG~\cite{HFAG} 
based on BLNP model~\cite{BLNP},
we obtain $f_B = 242^{+28}_{-27}\pm0.33$ \text MeV.
The obtained $f_B$ value is consistent with the unquenched lattice QCD calculation 
by HPQCD collaboration $f_B=216\pm22$ MeV~\cite{Gray:2005ad}.

\begin{figure}[bh]
    \begin{center}
	\includegraphics[width=0.3\textwidth]{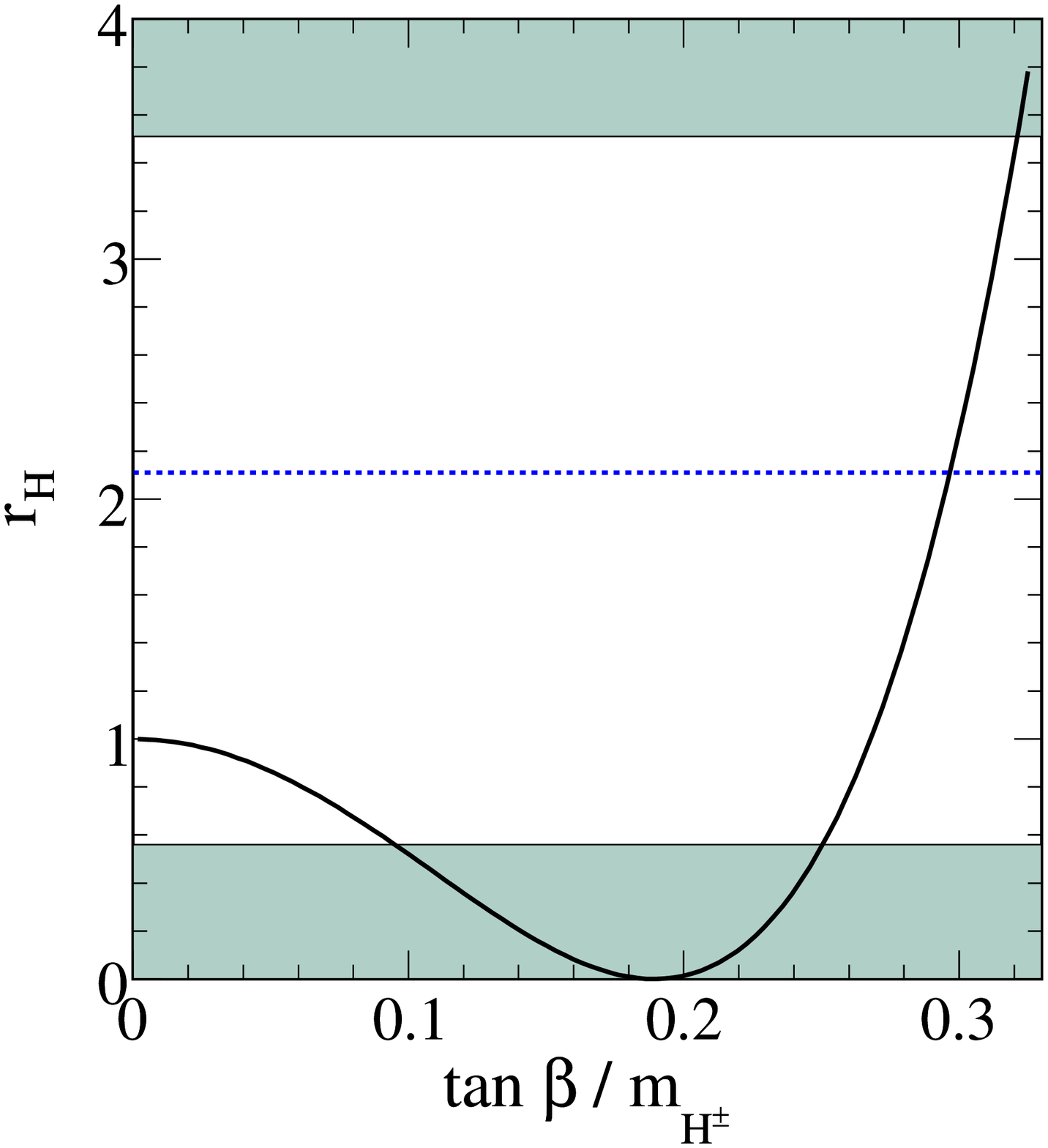}
	\includegraphics[width=0.3\textwidth]{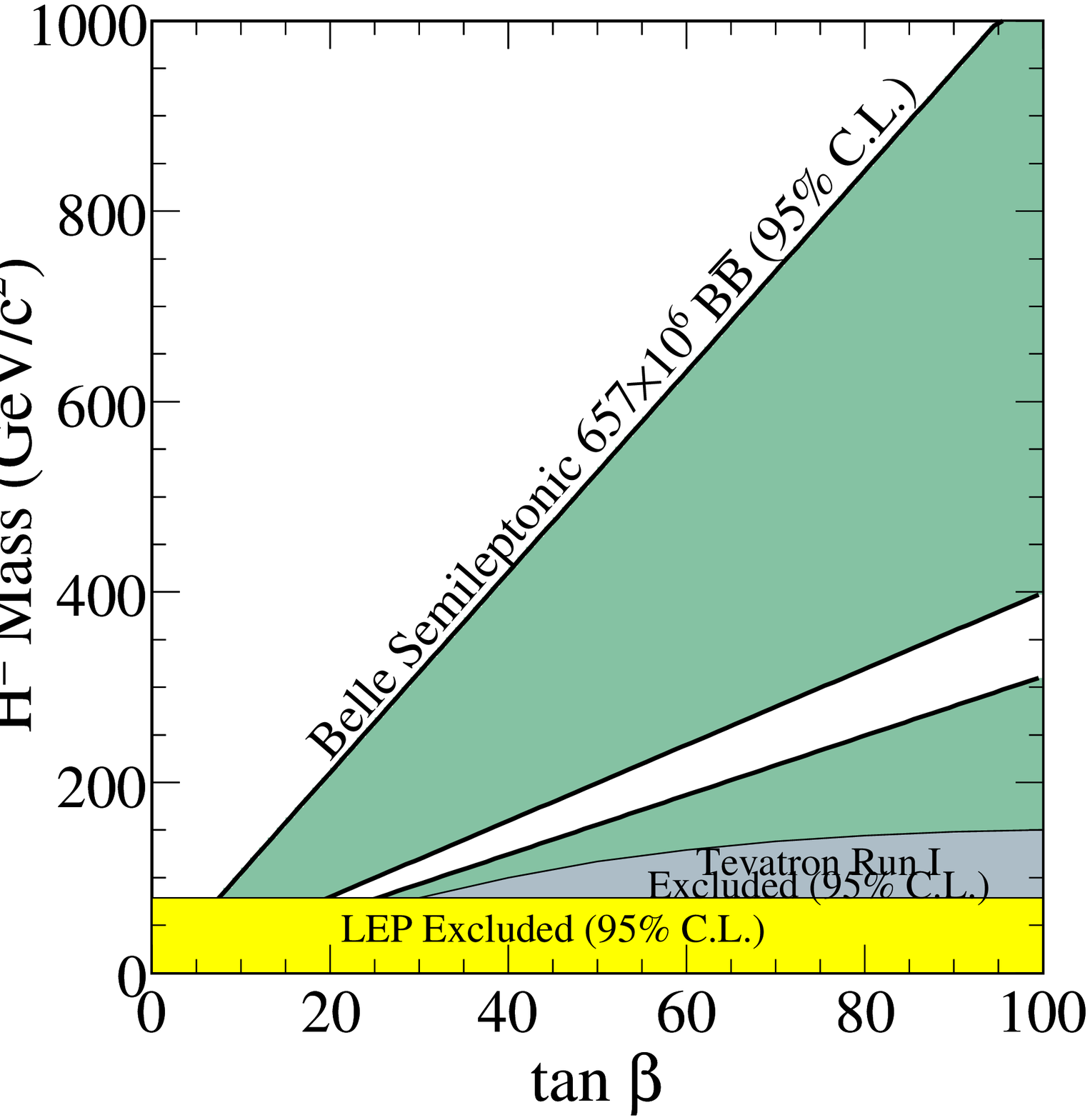}
	\caption{
	  Constraint on charged Higgs in $r_H$-$\tan\beta/m_H$ plane (left) and
	  $m_H$-$\tan\beta$ plane (right).
	  The black line in the right plot shows the expectation for $r_H$ 
	  as a function of $\tan\beta/m_H$ given by Eq.~\ref{eq:BR_higgs}.
	  The shaded areas indicate the excluded region with a confidence level of 95\%.
	  }
	\label{fig:chiggs}
    \end{center}
\end{figure}

The SM expectation for the branching fraction of $B\to\tau\nu$ from other experimental 
constraints is obtained by CKMfitter group to be 
${\cal B}(B^-\to\tau^-\overline{\nu}_\tau) = (0.78^{+0.09}_{-0.13})\times 10^{-4}$~\cite{CKMfitter2008}.
Comparing our result to it, we obtain $r_H = 2.11\pm0.75$,
where error include both statistical and systematic errors.
Constraint on charged Higgs is obtained using Eq.~\ref{eq:BR_higgs}.
Figure~\ref{fig:chiggs} shows the constraint on $\tan\beta$ and $m_H^\pm$.
The solid line in the left plot shows the expected $r_H$ 
as a function of $\tan\beta/m_H^{\pm}$ given by Eq.~\ref{eq:BR_higgs}.
The shaded areas are the excluded region with a confidence level of 95\%.

\section{Summary}
In summary, we have measured
the decay $B^{-}\rightarrow\tau^{-}\overline{\nu}$ with \bbbar pair events tagged by
semileptonic $B$ decays
using a data sample containing $657\times 10^6$ \bbbar pairs collected at the $\Upsilon(4S)$
resonance with the Belle detector at the KEKB asymmetric $e^{+}e^{-}$ collider. 
We measure the branching fraction to be
$(\brvalue\brstaterr(\text{stat})\brsysterr(\text{syst}))\times 10^{-4}$
with a significance of 3.8 standard deviations. 
We confirm the evidence reported in the previous Belle measurement
with \bbbar pair events tagged by hadronic $B$ decays.
The measured branching fraction is consistent with the SM expectation
from other experimental constraints.
The $B$ meson decay constant $f_B$ and constraint on charged Higgs are 
obtained using the measured branching fraction.


%

\begin{acknowledgments}
We thank the KEKB group for excellent operation of the
accelerator, the KEK cryogenics group for efficient solenoid
operations, and the KEK computer group and
the NII for valuable computing and SINET3 network
support.  We acknowledge support from MEXT and JSPS (Japan);
ARC and DEST (Australia); NSFC (China); 
DST (India); MOEHRD and KOSEF (Korea); 
KBN (Poland); MES and RFAAE (Russia); ARRS (Slovenia); SNSF (Switzerland); 
NSC and MOE (Taiwan); and DOE (USA).
\end{acknowledgments}


\begin{thebibliography}{99}
\bibitem{conjugate}
Throughout this paper, 
the inclusion of the charge conjugate mode decay is implied
unless otherwise stated.
\end{thebibliography}
\end{document}